\documentclass[aps,prb,twocolumn,showpacs,twoside]{revtex4}
\usepackage{graphicx,amssymb,amsmath,amsbsy,ifpdf,subfigure,dsfont,hyperref,ifthen}
\newboolean{submitToPRB} \setboolean{submitToPRB}{false}
\newcommand{\mycaption}[1]{\ifthenelse{\boolean{submitToPRB}}{\caption{(Color online) #1}}{\caption{#1}}}
\ifpdf \usepackage{color}
\hypersetup{pdfstartview=FitH,pdfauthor={Oleg CHALAEV and Daniel Loss <shalaev.oleg@unibas.ch>, Department of Physics, %and Astronomy, 
University of Basel, Klingelbergstrasse 82, Basel, CH-4056, Switzerland}}%
\else \usepackage[dvips]{color} \fi
\newcommand{\pf}{p_\mathrm F}
\newcommand{\be}{\begin{equation}}\newcommand{\ee}{\end{equation}}
\newcommand{\bes}{\begin{equation*}}\newcommand{\ees}{\end{equation*}}
\newcommand{\bea}{\begin{eqnarray}} \newcommand{\eea}{\end{eqnarray}}
\newcommand{\beas}{\begin{eqnarray*}} \newcommand{\eeas}{\end{eqnarray*}}
\newcommand{\GR}{G_{\mathrm R}}\newcommand{\GA}{G_{\mathrm A}}
\newcommand{\GRZ}{G_{\mathrm R}^{(0)}}\newcommand{\GAZ}{G_{\mathrm A}^{(0)}}\newcommand{\GRAZ}{G_{\mathrm R/A}^{(0)}}
\newcommand{\GRE}{G^E_{\mathrm R}}
\newcommand{\GAEw}{G^{E-\omega}_\mathrm A}
\newcommand{\hGR}{{\hat G}_{\mathrm R}}\newcommand{\hGA}{{\hat G}_{\mathrm A}}\newcommand{\hGRA}{{\hat G}_{\mathrm R/A}}
\newcommand{\Sp}{\mathop{\mathrm{Tr}}}\newcommand{\ud}{\mathrm{d}}

\newcommand{\hgr}{{\hat g}_{\mathrm r}}\newcommand{\hga}{{\hat g}_{\mathrm a}}\newcommand{\hgra}{{\hat g}_{\mathrm{r/a}}}

\newcommand{\cb}[1]{\colorbox{yellow}{#1}}
\newcommand{\EF}{E_\mathrm F}
\newcommand{\matr}[4]{\begin{pmatrix} #1&#2\\ #3&#4\end{pmatrix}}
\newcommand{\ipd}{\int\frac{\ud^2p}{(2\pi)^2}}
\newcommand{\ikd}{\int\frac{\ud^2k}{(2\pi)^2}}\newcommand{\iqd}{\int\frac{\ud^2q}{(2\pi)^2}}

\newcommand{\elib}[1]{\ifpdf \href{file://#1}{\includegraphics[height=.5cm]{dvd}}\fi}
\newcommand{\mybf}[1]{{\bf #1}}\renewcommand{\vec}{\mybf}
\newcommand{\ts}{{\tilde\sigma}} \newcommand{\bts}{\bar\sigma}
\newcommand{\sbarra}{\left|\vphantom{\Big[}\right.}

\renewcommand{\cb}{}
\let\labOrigRightarrow=\Rightarrow   \RequirePackage{marvosym}   \let\Rightarrow=\labOrigRightarrow
\begin{document}
\title{Anisotropic conductivity of disordered 2DEGs due to spin-orbit interactions}
\author{Oleg Chalaev} \author{Daniel Loss}
\affiliation{Department of Physics, University of Basel, Klingelbergstrasse 82, 4056 Basel, Switzerland}
%\date{August 27, 2007}
%\date{September 4, 2007}
%\date{September 29, 2007}
%\date{January 21, 2008}
\date{March 31, 2008}
\ifpdf
\hypersetup{pdfstartview=FitH,% <-- when opened in Acrobat Reader, starts in "page-screen-wide" mode
pdftitle={Anisotropic conductivity of a diffusive matter in the presence of Rashba and Dresselhaus spin-orbit coupling},%
pdfsubject={How Rashba and Dresselhaus spin-orbit interaction affects charge conductivity. Calculation using disorder-averaging diagrammatic technique},%
pdfauthor={Oleg CHALAEV <shalaev.oleg@unibas.ch> and Daniel Loss,
Department of Physics and Astronomy, University of Basel, Klingelbergstrasse 82, Basel, CH-4056, Switzerland},%
pdfkeywords={mesoscopics,disorder averaging,diffuson,diffusion approximation,Rashba,Dresselhaus,spin-orbit}}%
\fi
\begin{abstract}
We show that the disorder-averaged conductivity tensor of a disordered two-dimensional electron gas becomes anisotropic in the presence of both Rashba and Dresselhaus spin-orbit interactions
(SOI). This anisotropy is a mesoscopic effect and vanishes with vanishing charge dephasing
time. Using a diagrammatic approach including zero, one, and two-loop diagrams, we show that a consistent calculation needs to go beyond a Boltzmann equation approach.  In the absence of charge dephasing and for zero frequency, a
finite anisotropy $\sigma_{xy}\propto e^2/\pf lh$ arises even for infinitesimal SOI,
where $\pf$ is the Fermi momentum and $l$ is the mean free path of an electron.
\end{abstract}
%72.15.Rn, 71.55.Jv, 72.15.-v} % <-- put spaces in this list
% PACS:
% 72.10.-d 	Theory of electronic transport; scattering mechanisms
% 72.15.-v 	Electronic conduction in metals and alloys
%72.20.-i 	Conductivity phenomena in semiconductors and insulators 
% 73.23.-b 	Electronic transport in mesoscopic systems
% 73.20.Fz 	Weak or Anderson localization
%72.25.Dc 	Spin polarized transport in semiconductors
% 72.80.Ey 	III-V and II-VI semiconductors
% Metals / spin polarized transport in, 72.25.Ba
% (Anderson) Localization / conductivity in metals and alloys, 72.15.Rn
% Disordered solids / localization in, 71.55.Jv
% Metals / transport processes in, 72.15.-v
%
% Electron scattering /  elastic scattering, 34.80.Bm
% Anderson localization /  disordered solids, 71.23.An
% Localization /  weak, 72.15.Rn, 73.20.Fz
%
% Disordered solids / electrical conductivity, 72.80.Ng
% Amorphous metals and alloys / electrical and thermal conductivity, 72.15.Cz
%\pacs{72.10.-d, 73.63.-b, 72.15.Gd}
\maketitle
\section{Introduction\label{sec:intro}}
The interplay between spin and charge coherence effects in mesoscopic
semiconductors produces interesting transport phenomena~\cite{spintronics,Zutic,Awschalom:2007}. They are based on  the spin-orbit interaction (SOI),
such as Rashba~\cite{Rashba} or Dresselhaus~\cite{Dresselhaus} type, which establishes a coupling between orbital and spin degrees of freedom of the electron.

One major effect of SOI  on the conductivity  of a disordered semiconductor is the sign reversal of weak localization effects.
In this respect the influence of the SOI  is similar to that of a magnetic field:  it increases
the conductivity.  For Rashba and Dresselhaus SOI and in the presence of a magnetic field, this ``weak antilocalization'' effect
has been studied in a number of papers~\cite{Edelstein:95a,localSkvortsov,Zumbuhl,AleinerWL}.

In the absence of magnetic fields, the Rashba SOI  cannot violate isotropy of the energy spectrum.  In this case, the
conductivity tensor $\sigma_{\alpha\beta}$ is invariant under rotation of the coordinate system (CS) by $\pi/2$ and
simultaneous sign change of the SOI.
However, this sign is irrelevant for the conductivity (see Appendices), and thus
$\sigma_{xx}=\sigma_{yy}$ and $\sigma_{xy}=-\sigma_{yx}$. In addition, due to time reversal invariance,
$\sigma_{xy}=\sigma_{yx}$, so that $\sigma_{\alpha\beta}$ remains isotropic, i.e. $\sigma_{xy}=0$.
This reasoning is no longer valid if rotational invariance of the spectrum  is broken, e.g., by  a Zeeman term which
violates both time-reversal and rotational symmetries, leading to a finite anisotropy of the conductivity~\cite{Schwab}. 

A similar situation arises even without magnetic fields but when different types of SOI (such as Rashba and Dresselhaus) are present: 
the spectrum becomes anisotropic~\cite{SOIanCoDJ} so that one can expect anisotropy of the conductivity.
However, differently from before~\cite{Schwab}, a system with only SOI  remains
invariant under time reversal.  Nevertheless, we demonstrate below for a disordered two-dimensional
electron gas (2DEG) that breaking of the rotation symmetry of the spectrum alone 
leads to an anisotropic conductivity. 

Previous calculations
in such a system were based on the Boltzmann equation~\cite{SOIanCoDJ,JC:2006}, 
with the outcome~\cite{error} 
that the conductivity does get enhanced by the SOI, but remains isotropic even when both Rashba and Dresselhaus terms are present.

However, in the presence of phase coherence, both for charge and spin, the standard Boltzmann approach  is no longer sufficient.
Indeed, this approach correctly describes contributions to $\sigma_{\alpha\beta}$ of the
order of the Drude conductivity $\sigma_{\mathrm D}=e^2\pf l/2h$, but, as is well-known,  it already 'lacks accuracy' to 
describe weak localization corrections $\sim\sigma_{\mathrm D}\hbar/\pf l\ll\sigma_{\mathrm D}$
(here $\pf$ and $l$ are Fermi momentum and mean free path of the electrons, {\em resp.}). 
%
%To determine the anisotropy of the conductivity, we will see that
%even higher accuracy (i.e. one power more in $1/\pf l$) is needed.
We will see
that the required accuracy for finding the anisotropy of the
conductivity  is even  higher. 
%(i.e. one power more in $1/\pf l$).
%order in $1/\pf l$.

In the diagrammatic approach~\cite{AGD} used below, the leading contribution of a diagram to $\sigma_{\alpha\beta}$ is of the order
of $\sigma_{\mathrm D}/(\pf l/\hbar)^n$, where $n$ is the number of loops built by cooperon and diffuson lines in the diagram. (This is often called ``loop expansion''.)
Thus, the zeroth order is represented by two diagrams: the Drude ``bubble'' and the ``vertex correction'', which are often
referred to as ``zero loop approximation'' (ZLA)~\cite{aug04}, indicating  that these two diagrams have no loops made of cooperon and/or diffuson lines.
One can easily check that, in leading order,  these ZLA-diagrams are SOI-independent  (the ``bubble'' gives $\sigma_{\mathrm D}$, while the
``vertex correction'' vanishes).

The SOI-dependent contribution to $\sigma_{\alpha\beta}$, coming from the ZLA, is {\em isotropic} and on the order $\sigma_{\mathrm D}/(\pf l/\hbar)^2$. We will see
that this is of the same order as contributions from diagrams with two loops. 
Thus, a consistent calculation (i.e. a systematic expansion in powers of $\hbar/\pf l$) requires consideration of all the diagrams having zero, one, and two loops.
Below we demonstrate that the isotropic conductivity, obtained from the Boltzmann equation~\cite{JC:2006}, corresponds to the ZLA.  The inconsistency of the ZLA,
and thus of the Boltzmann equation, has been pointed out before~\cite{aug04} in the context of the spin-Hall effect.

Taking all relevant diagrams systematically into account, we find
that the {\em conductivity has finite anisotropic components} given by Eq. (\ref{finRe}), or,  when expressed in terms
of charge and spin dephasing times, by Eq. (\ref{anisRT}).
In the fully phase coherent limit,  a
finite anisotropy $\sigma_{xy}\propto e^2/2\pi\pf l$ exists even for infinitesimal SOI.

%\msc{Should we eliminate the following paragraph? (because it may look similar with the last paragraph of this Sec.) }
%We begin with  calculating the SOI-dependent contribution within the ZLA.
%Then, we demonstrate that this ZLA result is incomplete in a given order $\hbar/\pf l$, and higher order diagrams have to be considered.
%In the limit of small anisotropy of the spectrum (induced by SOI), the
%anisotropic conductivity $\sigma_{xy}$ is determined by  two-loop diagrams, which we then calculate explicitly.

%%Since the considered system is invariant under time reversal, $\sigma_{\alpha\beta}$ is symmetric and can thus be  diagonalized.
%%We will proceed with calculations in the coordinate system rotated by $\pi/4$ with respect to the original one,
%%where the 2D conductivity tensor is diagonal.

The paper is organized as follows. In Sec.~\ref{sec:Kubo}, we calculate matrix elements of the diffuson at zero frequency;
the result leads to the cancellation of the anomalous part of the velocity operator.
Then, in Sec.~\ref{sec:ZLA} we realize that the (most commonly used) zero loop approximation (ZLA) gives an isotropic contribution to the conductivity tensor,
similar to the result of the Boltzmann equation~\cite{JC:2006}. Then, we demonstrate that this ZLA result is incomplete in a given order $\hbar/\pf l$, and higher order diagrams have to be considered.
In Sec.~\ref{sec:hio} we obtain the general form of the expansion for the anisotropic part of the conductivity.
The main results are obtained in Sec.~\ref{sec:twoloops}, where we calculate the leading contribution to the anisotropy determined by two-loop diagrams and give estimates for real samples. 

\section{Kubo Formula and Vertex Renormalization\label{sec:Kubo}}

%Starting from the Kubo formual, we begin with  calculating the SOI-dependent contribution within the ZLA.
%Then, we demonstrate that this ZLA result is incomplete in a given order $\hbar/\pf l$, and higher order diagrams have to be considered.
%In the limit of small anisotropy of the spectrum (induced by SOI), the
%anisotropic conductivity $\sigma_{xy}$ is determined by  two-loop diagrams, which we then calculate explicitly.

Since the considered system is invariant under time reversal, $\sigma_{\alpha\beta}$ is symmetric and can thus be  diagonalized.
We will proceed with calculations in the CS rotated by $\pi/4$ with respect to the original one,
where the 2D conductivity tensor is diagonal.

In linear-response theory, the conductivity tensor is given by the Kubo-Greenwood formula:
\be \label{ofodc}
\sigma_{\alpha\beta}=\frac{e^2}h\overline{\Sp\left[{\hat v}_\alpha\hGR {\hat v}_\beta\hGA\right]},\quad\alpha,\beta=x,y,
\ee
where the overbar indicates averaging over the different disorder realizations,
${\hat v}_\alpha=\frac i\hbar[\hat H,r_\alpha]$ is a component of the velocity operator, and $\hGRA=[\EF-\hat H\pm i0]^{-1}$ with  $\EF$ being the Fermi energy
(the derivation of (\ref{ofodc}) is analogous to the one in~\cite{aug04}).
The Hamiltonian in our (rotated) CS reads
\be \label{Ham} \begin{split}
\hat H=&\frac{\hat p^2}{2m}+V_s+U(\vec r),\\
V_s=&(a-b){\ts}_1\hat p_y-(a+b){\ts}_2\hat p_x,
\end{split} \ee
where $a$ and $b$ are the amplitudes of Rashba and Dresselhaus SOI, ${\ts}_{1,2}$ are expressed in terms of Pauli matrices as
${\ts}_{1,2}=(\sigma_2\pm\sigma_1)/\sqrt2$, and $U(\vec r)$ is a short-range impurity potential, $\overline {U(\vec r)U(\vec r\,')}=\hbar^3(m\tau)^{-1}\delta(\vec
r-\vec r\,')$ (we use notations similar to~\cite{aug04}).
The conductivity (\ref{ofodc}) is not affected by a simultaneous sign-reversal of  $a$ and $b$;
thus, without loss of generality, we can assume that $a+b>0$.

In the absence of SOI, the disorder-averaged retarded Green function (GF) is given by~\cite{AGD}
\be \label{agf}
\GRZ=\left[G_{\mathrm A}^{(0)}\right]^\dag=\sigma_0\left[\EF-\frac{p^2}{2m}+\frac{i\hbar}{2\tau}\right]^{-1},
\ee
where $\sigma_0$ is the $2\times2$ unity matrix (due to spin degrees of freedom),
and $\tau$ is the mean time between collisions of an electron off impurities.
The presence of the SOI modifies (\ref{agf}) as follows:
\be \label{grSOI}
\GR=\GRZ\sum_{n\ge0}\left(V_s\GRZ\right)^n
=\left[{\left(\GRZ\right)^{-1}-V_s}\right]^{-1}.
\ee
%\cb{[Eq. (\ref{grSOI}) is the solution of a Dyson equation.]}

The disorder-averaging of (\ref{ofodc}) produces an infinite number of diagrams classified according to the number of loops composed by cooperon and diffuson
lines.  Each of these diagrams contains two velocity vertices, both~\cite{exceptZLA} being renormalized by the vertex correction~\cite{aug04},
% \be
% {\hat v}_\alpha=\frac i\hbar\left[\frac{{\hat p}^2}{2m}+V_s,r_\alpha\right],
% \ee
so that their ``anomalous'' (i.e., SOI-dependent) part cancels:
\be\label{renaVer}\begin{split}
\hat{\vec v}=&\frac{\hat{\vec p}}m+\begin{pmatrix}-(a+b){\ts}_2\\(a-b){\ts}_1\end{pmatrix},\\
{\hat{\tilde v}}_\alpha=&{\hat v}_\alpha+\sum_{\gamma=1}^3{\ts}_\gamma D^{\gamma\gamma}\Sp_{\vec p}\left[{\ts}_\gamma\GR(\vec p){\hat v}_\alpha\GA(\vec p)\right]
=\frac{{\hat p}_\alpha}m,
\end{split}\ee
where
\be\begin{split}
(&D^{11},D^{22},D^{33})=\\ =&\frac\hbar{m\tau}\left(1+\frac{1+K}{(x_a+x_b)^2},1+\frac{1+K}{(x_a-x_b)^2},\frac K{K-1}\right)
\end{split}\ee
are the components of the diffuson at zero momentum,
$K=\sqrt{[1+(x_a+x_b)^2][1+(x_a-x_b)^2]}$, and we have introduced dimensionless Rashba and Dresselhaus amplitudes,
$x_a=2\pf a\tau/\hbar$ and $x_b=2\pf b\tau/\hbar$. [Note that (\ref{renaVer}) is not exact at finite frequency.]

\section{The zero-loop approximation\label{sec:ZLA}}
In the ZLA, the conductivity  is given by two diagrams -- the ``bubble'' and the ``vertex correction''~\cite{aug04},
which, when being summed, result in the following SOI-dependent correction to the conductivity \cite{foot}:
\be \begin{split}
\delta\sigma_{\alpha\beta}=
\frac{e^2}h\Sp_{\vec p}\big[{\hat{\tilde v}}_\alpha\GR(\vec p){\hat v}_\beta\GA(\vec p)&\\
-{\hat{\tilde v}}_\alpha\GRZ(\vec p){\hat v}_\beta\GAZ(\vec p)\big]
=\frac{e^2}h&\frac{x_a^2+x_b^2}{2\pf l/\hbar}\matr1001,\label{ZLAres}
\end{split} \ee
where $\GRAZ$ is given in (\ref{agf}) and $\Sp_{\vec p}\equiv\ipd$.
The result (\ref{ZLAres}) has been confirmed (up to a missing factor $4$) in~\cite{JC:2006} by using the Boltzmann equation approach;
so one can conclude that ZLA corresponds to the result of the Boltzmann equation with the simplest (and most common) form of the collision integral.
However, since (\ref{ZLAres}) is of the order of $\sigma_{\mathrm D}/(\pf l/\hbar)^2\ll\sigma_{\mathrm D}$, the calculation in the ZLA (as well as the Boltzmann
equation~\cite{JC:2006}) is incomplete.

To re-enforce this point, let us consider the special case when $a=\pm b$.  Then, all four operators under the $\Sp$ in (\ref{ofodc}) can be diagonalized by a
momentum-independent unitary transformation, and one can see that the SOI-dependent correction to the conductivity vanishes already before the disorder
averaging (see Appendix~\ref{appB}). This is just further evidence that the expression (\ref{ZLAres}) is incomplete; other contributions must be taken into account to cancel it when $a=\pm b$.

\section{Contribution of higher orders\label{sec:hio}}
According to the loop expansion~\cite{aug04}, diagrams having one (weak localization~\cite{Edelstein:95a,localSkvortsov}) and two loops may produce contributions
to $\sigma_{\alpha\beta}$ of the same or even larger order than (\ref{ZLAres}).  There exist one diagram with one loop and nine diagrams with two loops.

{}From now on we assume that  the SOI and the spectrum anisotropy due to it are small:
\be \label{assump}
x=\sqrt{x_a^2+x_b^2}\ll1,\ \delta=\frac{2ab}{a^2+b^2},\ |\delta|\ll1.
\ee
We have chosen $x$ and $\delta$ as expansion parameters because they provide a uniform expansion for the conductivity~(\ref{pervoryadDSi}).

%\msc{here I've erased piece of text}
The anisotropic part of the conductivity tensor is given by an expansion in inverse powers of $\pf l/\hbar\gg1$:
\be \label{pervoryadDSi}
\sigma_{xx}-\sigma_{yy}=2\frac{e^2}h\sum_{r\ge0}\frac1{(\pf l/\hbar)^{r-1}}\sum_{m,n\ge0}S_{mn}^rx^m\delta^{2n+1},
\ee
where we used~(\ref{deltaSigmaxxyy}).

By calculating the conductivity in the ZLA we have checked that  $S_{mn}^0=0$ $\forall m,n$.
From the properties of the weak localization (WL)
diagram (which is the only one that could contribute to $S_{00}^1$), one can easily show that $S_{00}^1=0$ as well.
Thus, in the limit when $\pf l\delta^2/\hbar<1$, the leading anisotropic contribution   is given by the term $\propto S_{00}^2$ in the
expansion (\ref{pervoryadDSi}), which we calculate below.

\section{Diagrams with two loops\label{sec:twoloops}}
\begin{figure}
% \subfigure[]{\label{fdivtoPor:b}%
% \begin{minipage}{.15\textwidth}\resizebox{\textwidth}{!}{\rotatebox{90}{\input{cc_SOI-b.ins.tex}}}\end{minipage}}
% \subfigure[]{\label{fdivtoPor:c}%
% \begin{minipage}{.15\textwidth}\resizebox{\textwidth}{!}{\rotatebox{90}{\input{cc_SOI-c.ins.tex}}}\end{minipage}}
% \subfigure[]{\label{fdivtoPor:d}%
% \begin{minipage}{.15\textwidth}\resizebox{\textwidth}{!}{\rotatebox{90}{\input{cc_SOI-d.ins.tex}}}\end{minipage}}
\includegraphics{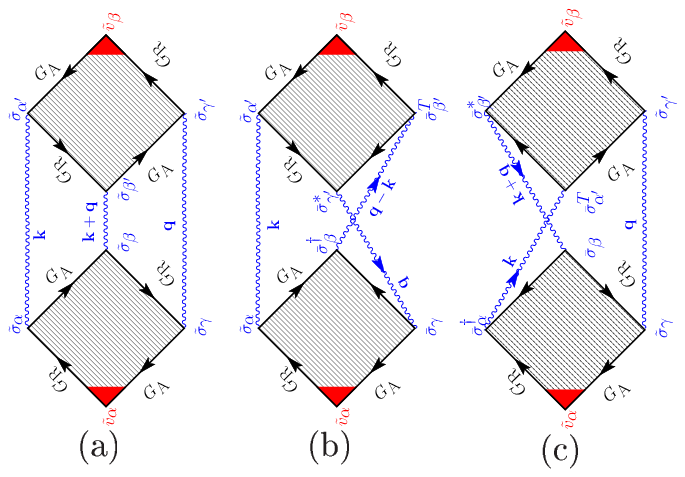}
\mycaption{Three relevant two-loop diagrams (out of nine) for the conductivity tensor $\sigma_{\alpha\beta}$.  The matrix vertices $\ts$ and $\bts$ are written
assuming clockwise (in lower Hikami boxes) and counterclockwise (in upper Hikami boxes) direction of writing traces ($\Sp$).  Each diagram contains
small-momentum singularities, which cancel each other in accordance with the theorem from~\cite{woelfe}.
\label{fdivtoPor}} \end{figure}

We have checked that, like the ZLA, the WL diagram does not contribute to $S_{00}^2$. The same is true for 6 (out of 9) two-loop diagrams.
The remaining 3 diagrams, which we calculate below, are depicted in Fig.~\ref{fdivtoPor}.
In addition to the figure caption, some comments are necessary. 
Straight bold lines represent averaged retarded and advanced GFs, given by (\ref{grSOI}) and its Hermitian conjugate.
The squares built of GFs are Hikami boxes~\cite{localGLK}.
A Hikami box~(HB) is given by three contributions, cf.~\cite{aug04}: one empty box and two boxes with an impurity line inside.
Wavy lines with (without) arrows represent cooperons (diffusons).

Using the identity $\sum_{\alpha=0}^3\ts^{s_1s_3}_\alpha{\ts}^{s_4s_2}_\alpha=2\delta_{s_1s_2}\delta_{s_3s_4}$, we
write the diffuson in the basis of matrices $\ts$ as
$D_{\vec q}=\frac\hbar{2m\tau}\left[\mathds1-{\tilde X}_D(\vec q\,)\right]^{-1}$, where $\mathds1$ is a $4\times4$ unity matrix, and
\be\label{defXabDC}
{\tilde X}_D^{\alpha\beta}(\vec q\,)=\frac\hbar{2m\tau}\Sp_{\vec p}[\ts_\alpha\GRE(\vec p\,)\ts_\beta\GAEw(\vec p-\vec q\,)].
\ee
For cooperon lines in diagrams \ref{fdivtoPor}b %\ref{fdivtoPor:c}
and \ref{fdivtoPor}c, %\ref{fdivtoPor:d},
we use the expression
$C_{\vec q}=\frac\hbar{2m\tau}\left[\mathds1-{{\bar X}}_C(\vec q\,)\right]^{-1}$, so that the cooperon series starts with \emph{one} disorder line.
Thus each diagram in Fig.~\ref{fdivtoPor} contains an extra contribution, belonging to the WL diagram.
However, this contribution can affect $S_{mn}^r$ in (\ref{pervoryadDSi}) only for $m\ge1$, so we neglect it.
To make the similarity between diffuson and  cooperon more evident, we use a different matrix basis for the cooperon:
\be
{{\bar X}}_C^{\alpha\beta}(\vec q\,)=\frac\hbar{2m\tau} \label{altXc}
\Sp_{\vec p}\{{\bar\sigma}_\alpha\GRE(\vec p\,){\bar\sigma}^\dag_\beta[\GAEw(\vec q-\vec p\,)]^T\},
\ee
where ${\bts}_\alpha=\sigma_2{\ts}_\alpha$, and
$\sum_{\alpha=0}^3{{\bar\sigma}}^{s_1s_3}_\alpha\left({{\bar\sigma}}_\alpha^\dagger\right)^{s_4s_2}=2\delta_{s_1s_2}\delta_{s_3s_4}$.
Because of the time-reversal symmetry, $\sigma_2G_{\mathrm A}^T(-\vec p\,)\sigma_2=\GA(\vec p\,)$, so that ${{\bar X}}_C(\vec q\,)={\tilde X}_D(\vec q\,)$ and
$C_{\vec q}=D_{\vec q}$. Thus the expressions for cooperon/diffuson lines are the same for all diagrams in Fig.~\ref{fdivtoPor}, and the same is true for the
lower HBs $L_{\alpha\beta\gamma}$. As a consequence, the sum of the three diagrams is equal to the diagram in Fig.~\ref{fdivtoPor}a %\ref{fdivtoPor:b}
with a
``renormalized'' upper HB $R_{\alpha'\beta'\gamma'}$ (given by the sum of upper HBs of all three diagrams), see the expression for
$\delta\sigma_{xx}^{\mathrm{II}}$ in (\ref{glVPsl}).

Due to (\ref{deltaSigmaxxyy}) the anisotropy of the conductivity is given by
$\sigma_{xx}(a,b)-\sigma_{yy}(a,b)=\sigma_{xx}(a,b)-\sigma_{xx}(a,-b)$, so it suffices to calculate only the SOI-dependent part of $\sigma_{xx}$.
The total contribution of the diagrams in Fig.~\ref{fdivtoPor} is an integral from the product of two HBs ($L$ and
$R$) and three diffusons ($D$):
\be \begin{split}
\delta\sigma_{xx}^{\mathrm{II}}=\frac{e^2}h\ikd\iqd\sum_{\alpha,\beta,\gamma=0}^3\\ \sum_{\alpha',\beta',\gamma'=0}^3L_{\alpha\beta\gamma}
D^{\alpha\alpha'}_{\vec k}D^{\beta'\beta}_{\vec k+\vec q}D^{\gamma\gamma'}_{\vec q}R_{\alpha'\beta'\gamma'}.  \label{glVPsl}
\end{split} \ee
First, we evaluate (\ref{glVPsl}) for zero frequency $\omega=0$.
In the absence of SOI, the HBs are given by
\be \begin{split}
L^{\mathrm 0}_{\alpha\beta\gamma}=-2i\frac{\pf l\tau^3}{\hbar^4}(k_x+q_x)\Sp[\ts_\alpha\ts_\beta\ts_\gamma],\\
R^0_{\alpha\beta\gamma}=
-4i\frac{\pf l\tau^3}{\hbar^4}\big\{(1-\delta_{\beta0})(k_x\delta_{\alpha0}\delta_{\beta\gamma}+q_x\delta_{\gamma0}\delta_{\alpha\beta})+\\
(k_x+q_x)(1-\delta_{\alpha0})\big[\delta_{\beta0}\delta_{\alpha\gamma}+i\epsilon_{\alpha\beta\gamma}(1-\delta_{\beta0})(1-\delta_{\gamma0})\big]\big\},
\end{split} \ee
where $\epsilon_{\alpha\beta\gamma}$ is the Levi-Civit\`a tensor. The SOI-dependent part of the HBs is shown in Tab.~\ref{tab:renHBslSOI}.
Due to SOI, additional nonzero matrix elements appear, $L^{\mathrm s}_{\alpha\beta\gamma}$ and $R^{\mathrm s}_{\alpha'\beta'\gamma'}$,
so that $L_{\alpha\beta\gamma}=L^{\mathrm 0}_{\alpha\beta\gamma}+L^{\mathrm s}_{\alpha\beta\gamma}$ and $R_{\alpha'\beta'\gamma'}=R^0_{\alpha'\beta'\gamma'}+R^{\mathrm s}_{\alpha'\beta'\gamma'}$, see Tab.~\ref{tab:renHBslSOI}.

\begin{table}\begin{center}\begin{tabular}{
|                                c  |  c |  c | c  |c    | c  |c    |c    |  c|}\hline
$\alpha$                            &$0$ &$0$ &$1$ &$1$  & $2$&  $2$&$3$  &$3$ \tabularnewline\hline
$\beta$   			    &$1$ &$3$ &$1$ &$3$  & $1$&  $3$&$1$  &$3$ \tabularnewline\hline
$\gamma$  		            &$3$ &$1$ &$2$ &$0$  & $1$&  $3$&$0$  &$2$ \tabularnewline\hline
$\frac{L^{\mathrm s}_{\alpha\beta\gamma}\hbar^3}{2\pf\tau^3(x_a+x_b)}$
                                    &$-1$&$1$  &$i$&$1$  & $-i$&$-i$&$-1$&$i$\tabularnewline\hline
$\frac{R^{\mathrm s}_{\alpha\beta\gamma}\hbar^3}{2\pf\tau^3(x_a+x_b)}$
                                    &$0$ &$0$ &$-2i$&$0$& $2i$ &$2i$ &$0$ &$-2i$ \tabularnewline\hline
\end{tabular}
\end{center}\caption{The first order SOI-correction to the Hikami boxes ($L^{\mathrm s}_{\alpha\beta\gamma}$ and $R^{\mathrm s}_{\alpha'\beta'\gamma'}$).
Other matrix elements vanish. Each of three upper Hikami boxes has non-zero elements with $\alpha=0$ or $\gamma=0$, so that each diagram diverges at $\vec
k\to0$ or at $\vec q\to0$; however, these elements cancel each other (note the zeroes in the last row), so that the complete expression for the
conductivity converges.\label{tab:renHBslSOI}}
\end{table}
% Like the HBs, ${\tilde X}_D$ defined in (\ref{defXabDC}) can be split into a SOI-independent part
% ${\tilde X}_D^{\mathrm 0}(\vec q\,)=\left(1+l^2q^2/2\right)\mathds1$ and the correction ${\tilde X}_D^{\mathrm s}$:
Like the HBs, ${\tilde X}_D$ defined in (\ref{defXabDC}) can be split into a 
main part and a SOI-dependent correction, which
(for $\omega=0$) are given by
${\tilde X}_D^{\mathrm 0}(\vec q\,)=\left(1-l^2q^2/2\hbar^2\right)\mathds1$ and
\bea\label{XdRD}
\lefteqn{{\tilde X}_D^{\mathrm s}(\vec q\,)=} \\ & & \nonumber
\begin{pmatrix}
0&0&0&0\\
0&\frac{(x_a+x_b)^2}2     &0           &-i\frac{q_xl}\hbar(x_a+x_b)\\
0&0     &\frac{(x_a-x_b)^2}2           &-i\frac{q_yl}\hbar(x_a-x_b)\\
0&i\frac{q_xl}\hbar(x_a+x_b) &  i\frac{q_yl}\hbar(x_a-x_b)     & x_a^2+x_b^2   \end{pmatrix}.
\eea
Hence the diffusons in (\ref{glVPsl}) are given by
\be\label{vyddi}
D_{\vec q}=\frac\hbar{2m\tau}\left[\mathds1-{\tilde X}_D^{\mathrm 0}(\vec q\,)-{\tilde X}_D^{\mathrm s}(\vec q\,)\right]^{-1},
\ee
where ${\tilde X}_D^{\mathrm s}$ is written in~(\ref{XdRD}).

Two angular integrations in (\ref{glVPsl}) were performed  analytically using {\tt maxima}~\cite{maxima} 
(the result is too lengthy to be presented here), while the integrations over $ k$, $ q$ were done numerically, using %procedures form the
{\tt quadpack}~\cite{quadpack}. 

Finally, we obtain the following result for the anisotropy of the conductivity at $\omega=0$,
\be\label{finRe} \begin{split}
\delta\sigma=2\delta\sigma_{xx}^{\mathrm{II}}=\sigma_{xx}-\sigma_{yy}=2S^2_{00}\frac{2x_ax_b}{x_a^2+x_b^2}\frac{e^2}{2\pi}\frac1{\pf l},\\
S^2_{00}=-5.6\times10^{-3},\quad 2x_ax_b\ll x_a^2+x_b^2\ll1.
\end{split}\ee

The result (\ref{finRe}) has been obtained in the  CS rotated by $\pi/4$, where the sum of Rashba and Dresselhaus terms are
given by $V_s$ in (\ref{Ham}). In the original coordinates, the conductivity has equal diagonal elements, and equal non-zero off-diagonal elements
$\sigma_{xy}=\sigma_{yx}=\delta\sigma/2$.

The correction (\ref{finRe}) depends non-analytically on the SOI in the vicinity of $x_a^2+x_b^2=0$, so that at $\omega=0$ even an infinitesimally small SOI can
produce a finite anisotropy. This is not entirely surprising, since a similar non-analyticity is also seen to emerge in the weak-localization correction with
Rashba SOI only and for vanishing dephasing (see, e.g., Eq. (17) in Ref.~\cite{localSkvortsov}).

The analyticity is restored at finite frequency:
\be \label{hochFR} \begin{split}
\delta\sigma=-2\cdot0.25\cdot\frac{-2i\omega\tau\cdot2x_ax_b}{(x_a^2+x_b^2-2i\omega\tau)^2}\frac{e^2}{2\pi}\frac1{\pf l},\\
2x_ax_b\ll x_a^2+x_b^2\ll\omega\tau\ll1.
\end{split}\ee
To account for charge dephasing effects, we follow standard procedure~\cite{Altshuler:82} and substitute
$-i\omega\tau\to\tau/\tau_\phi$, with
$\tau_\phi$ the charge dephasing time (e.g., due to
electron-electron interactions). In addition, one can express the SOI parameters in terms of spin dephasing times
of Dyakonov-Perel type~\cite{spinRelax}, allowing us to rewrite (\ref{finRe}), (\ref{hochFR}) as
\be\label{anisRT}
\delta\sigma=\left\{\aligned
5.6\times10^{-3}\cdot\frac{\tau_--\tau_+}{\tau_-+\tau_+}\frac{e^2}{2\pi}\frac1{\EF\tau},\quad\tau_\pm\ll\tau_\phi,\\
0.13\cdot\left(\frac{\tau_\phi}{\tau_+}-\frac{\tau_\phi}{\tau_-}\right)\frac{e^2}{2\pi}\frac1{\EF\tau},\quad\tau_\phi\ll\tau_\pm,
\endaligned \right.
\ee
where\cite{spinRelax} $2\tau/\tau_\pm=(x_a\mp x_b)^2$. {In a diffusive semiconductor (e.g., GaAs) for $\pf l/\hbar=5$} we estimate
$\delta\sigma\lesssim5\cdot10^{-4}\sigma_D$ in case $\tau_\pm\ll\tau_\phi$ and
$\delta\sigma\lesssim10^{-2}\sigma_D$ in the opposite case $\tau_\pm\gg\tau_\phi$.

Thus, in a system with finite dephasing time $\tau_\phi$, the anisotropic conductivity (\ref{anisRT}) is analytic in $\tau_\pm$.
However, in a fully phase-coherent system with $\tau_\phi=\infty$ (which is commonly predicted at zero temperature),
the conductivity tensor depends non-analytically on the SOI.

\section{Conclusion}
The combination of Rashba and Dresselhaus  SOI leads to anisotropy in the energy spectrum, which, in turn,
leads to anisotropy in the conductivity of a disordered 2DEG.
This anisotropy, being due to phase coherence of charge and spin, comes from two-loop diagrams in the perturbative approach, while it is 
absent in the Boltzmann equation approach. In the 
limit of full phase coherence, the system  ``jumps'' from isotropic to anisotropic state for infinitesimally small SOIs.

We are grateful to B.~Altshuler, M.~Duckheim,  A.~Khaetskii, and E.~Rashba for helpful discussions.
We acknowledge financial support from the Swiss NSF and the NCCR Nanoscience.

%\msc{start of changes}

\appendix
\section{Conductivity for pure-Rashba  or pure-Dresselhaus SOI \label{pureRD}}
Using the time-reversal invariance we argued in the Introduction (Sec.~\ref{sec:intro}) that the conductivity tensor is isotropic in case of a pure-Rashba SOI.
Here we demonstrate this isotropy for both pure-Rashba and pure-Dresselhaus cases using
only rotation symmetry of the disorder-free part of the Hamiltonian.

The conductivity tensor in the rotated CS is connected with its value in the original CS through the transformation
\be \label{preobr}
\sigma_{\alpha\beta}\to(R_\phi\sigma R_\phi^{-1})_{\alpha\beta},
\ee
where $R_\phi$ is a 3D-rotation matrix in the $(x,y)$-plane by %an \emph{arbitrary}
angle $\phi$.
In this section, we prove that when $a=0$ or $b=0$, the conductivity tensor is invariant with respect to rotations in the $(x,y)$-plane;
together with the requirement (\ref{preobr}) this means that $\sigma_{\alpha\beta}$ is isotropic.

We define matrices
\be
T=\frac1{\sqrt2}\begin{pmatrix}0&1+i\cr1-i&0\end{pmatrix},\quad
C=\begin{pmatrix}0 & -1 &  0 \cr  -1 & 0 & 0 \cr 0 & 0 & -1\end{pmatrix}.
\ee
The Hamiltonian (\ref{Ham}) can be rewritten as
\be
\hat H=\frac{{\hat p}^2}{2m}+aV_R+bV_D+U(\vec r),
\ee
where (in the original -unrotated- CS)
\be
V_R={\vec e}_z\cdot\left[\boldsymbol\sigma\times{\hat{\vec p}}\right],\quad V_D={\vec e}_z\cdot\left[C\boldsymbol\sigma\times{\hat{\vec p}}\right],
\ee
$\boldsymbol\sigma\equiv(\sigma_1,\sigma_2,\sigma_3)$, and ${\vec e}_z$ is the unit vector along the $z$-axis.
%where brackets denote the mixed product (oriented volume).
A rotation of the CS corresponds to the transformation of coordinates, momenta and spins according to
\be\label{stRot}
(\sigma_0,\boldsymbol\sigma)\to(\sigma_0,R_\phi\boldsymbol\sigma),\quad
\hat{\vec p}\to R_\phi\hat{\vec p},\quad
\vec r\to R_\phi\vec r.
\ee
We note that $V_R$ is invariant under the transformation (\ref{stRot}):
\be \label{invRashba}
%\forall\phi\quad 
V_R={\vec e}_z\cdot\left[R_\phi\boldsymbol\sigma\times R_\phi{\hat{\vec p}}\right]=\begin{pmatrix}0&{\hat p}_y+i{\hat p}_x\cr {\hat p}_y-i{\hat p}_x&0\end{pmatrix}.
\ee
The same invariance holds for
% the purely-Rashba velocity operator: ${\hat v}_\alpha^{Ra}=\frac i\hbar\left[\frac{p^2}{2m}+aV_R,r_\alpha\right]$, as well as for
the (disorder-averaged) pure-Rashba Green functions $G_{\mathrm R/A}^{Ra}$. Thus, the contribution to the conductivity tensor from the (simplest) bubble
diagram~\cite{aug04}
\be\begin{split}\label{RAI}
\sigma^{Ra}_{\alpha\beta}=&\frac{e^2}h\Sp_{\vec p}\Sp_{\mathrm{spin}}\left[{\hat v}_\alpha^{Ra}G_{\mathrm R}^{Ra}{\hat v}_\beta^{Ra}G_{\mathrm A}^{Ra}\right],\\
{\hat v}_\alpha^{Ra}=&\frac i\hbar\left[\frac{{\hat p}^2}{2m}+aV_R,r_\alpha\right]
\end{split}\ee
is invariant under the rotation (\ref{stRot}), and hence isotropic [due to \ref{preobr})] in a pure-Rashba SOI system.
As an example of a higher-loop correction, let us consider the two-loop contribution (\ref{glVPsl}). The symmetry transformation (\ref{stRot}) will affect
(\ref{glVPsl}) twofold:
(i) as a momentum rotation $\vec k,\vec q\to R_\phi\vec k,R_\phi\vec q$ and (ii) through a different set of Pauli matrices
$(\sigma_0,\boldsymbol\sigma')=(\sigma_0,R_\phi\boldsymbol\sigma)$ in the calculation of diffusons and Hikami boxes.
The rotation of momenta can be absorbed by a variable change in the integration.
%As for the change in the set of Pauli matrices, \cb{it seems obvious} that
Then,
the result of the calculation does not depend on the choice of Pauli matrices,
once they obey standard commutation relations. Since the new set $(\sigma_0,\boldsymbol\sigma')$ is connected through a unitary transformation with
the standard one, $(\sigma_0,\boldsymbol\sigma)$ the commutation relations are preserved so that  (\ref{glVPsl}) is invariant under the transformation (\ref{stRot}).
Thus, the two-loop contribution (\ref{glVPsl}) does not change when the CS is rotated in the $(x,y)$-plane.
Analogously, the contribution from any other diagram is invariant under the transformation (\ref{stRot}) in case when $b=0$, so that [due to \ref{preobr})]
the pure-Rashba conductivity tensor is isotropic.

Let us now demonstrate that also in case of pure-Dresselhaus SOI the conductivity is isotropic.
The Dresselhaus SOI can  be transformed into Rashba SOI by a unitary transformation: $V_D=T^\dag V_RT$;
the same is true for the pure-Dresselhaus velocity operator: ${\hat v}_\alpha^{Db}=\frac i\hbar\left[\frac{{\hat p}^2}{2m}+bV_D,r_\alpha\right]=T^\dag {\hat v}_\alpha^{Rb}T$.
So, one can transform the pure-Dresselhaus conductivity tensor into the pure-Rashba one:
\be \begin{split}\label{Dpoh}
\sigma^{Db}_{\alpha\beta}=&\frac{e^2}h\Sp\left[{\hat v}_\alpha^{Db}G_{\mathrm R}^{Db}{\hat v}_\beta^{Db}G_{\mathrm A}^{Db}\right]\\
=&\frac{e^2}h\Sp\left[T^\dag {\hat v}_\alpha^{Rb}TT^\dag G_{\mathrm R}^{Rb}TT^\dag {\hat v}_\beta^{Rb}TT^\dag G_{\mathrm A}^{Rb}T\right],
%=\sigma^{Rb}_{\alpha\beta},
\end{split} \ee
so that $\sigma^{Ra}_{\alpha\beta}=\sigma^{Da}_{\alpha\beta}$.
%Like in case of purely-Rashba SOI (\ref{RAI}),
With  (\ref{Dpoh}) we proved this statement for the (simplest) bubble diagram~\cite{aug04};
it can be generalized to higher-loop corrections analogously to the pure-Rashba case [see (\ref{RAI}) above].
We conclude that the \emph{disorder-averaged} conductivity tensor is isotropic in a system with either pure Rashba or pure Dresselhaus  SOI.
%(This argumentation is not valid, if the time-reversal symmetry is broken.)

\section{Conductivity in case  when $a=\pm b$\label{appB}}
The case, when the moduli of Rashba and Dresselhaus SOI amplitudes are equal, is  special~\cite{schliemann03:146801,schliemann03:165311,aug04}.
%In the special case $a=\pm b$,
Both Hamiltonian and velocity operators can then be diagonalized in spin space by unitary transformations: for $a=b$,
\cb{up to an irrelevant constant (in the rotated CS)}
\be
U_1\hat HU_1^\dag=\frac m2\left[\left(\frac{{\hat p}_x}m-2a\sigma_3\right)^2+\frac{{\hat p}_y^2}{m^2}\right]+U(\vec r),
\ee
and
\be
U_1\hat{\vec v}U_1^\dag=\begin{pmatrix} \frac{{\hat p}_x}m-2a\sigma_3\\ \frac{{\hat p}_y}m \end{pmatrix},
\ee
where the transformation matrix is given by
\be \label{yuone}
U_1=(U_1^\dag)^{-1}=\frac1{\sqrt2}\matr1{-e^{i\pi/4}}1{e^{i\pi/4}}.
\ee
For $a=-b$,
\be
U_2\hat HU_2^\dag=\frac m2\left[\left(\frac{{\hat p}_y}m+2b\sigma_3\right)^2+\frac{{\hat p}_x^2}{m^2}\right]+U(\vec r),
\ee
and
\be
U_2\hat{\vec v}U_2^\dag=\begin{pmatrix} \frac{{\hat p}_x}m\\ \frac{{\hat p}_y}m+2b\sigma_3\end{pmatrix},
\ee
with
\be \label{yutwo}
U_2=(U_2^\dag)^{-1}=\frac1{\sqrt2}\matr1{-e^{-i\pi/4}}1{e^{-i\pi/4}}.
\ee
Let us denote the (unaveraged) Green functions without SOI ($a=b=0$) as $\hgra$.

%Using (\ref{aEQbup}), we conclude that the conductivity tensor for $a=\pm b$ is the same, as without SOI. In fact,
With $U_{1,2}$ from (\ref{yuone}) and (\ref{yutwo}) we diagonalize all operators under the trace  $\Sp$ in the conductivity expressions below.
Then the spin part of $\Sp\equiv\Sp_{\mathrm{spin}}\Sp_{\vec p}$ splits into two terms. For $a=b$, 
\be \label{parpmbN}\begin{split}
&a=b\Rightarrow\Sp_{\vec p}\Sp_{\mathrm{spin}}\left[{\hat v}_x\hGR{\hat v}_x\hGA \right]\\
&=\Sp_{\vec p}\Sp_{\mathrm{spin}}\left[U_1{\hat v}_xU_1^\dag U_1\hGR U_1^\dag U_1{\hat v}_xU_1^\dag U_1\hGA U_1^\dag\right]\\
&=\Sp_{\vec p}\left\{
\left[\frac{{\hat p}_x}m\hgr\frac{{\hat p}_x}m\hga\right]\sbarra_{{\hat p}_x\to{\hat p}_x-2ma}+
\left[\frac{{\hat p}_x}m\hgr\frac{{\hat p}_x}m\hga\right]\sbarra_{{\hat p}_x\to{\hat p}_x+2ma}\right\}\\
&=2\Sp_{\vec p}\left[\frac{{\hat p}_x}m\hgr\frac{{\hat p}_x}m\hga\right]=2\Sp_{\vec p}\left[\frac{{\hat p}_y}m\hgr\frac{{\hat p}_y}m\hga\right],\\
\end{split}\ee
and for $a=-b$, 
\be\begin{split}
&a=-b\Rightarrow
\Sp_{\vec p}\Sp_{\mathrm{spin}}\left[{\hat v}_y\hGR{\hat v}_y\hGA \right]\\
&=\Sp_{\vec p}\Sp_{\mathrm{spin}}\left[U_2{\hat v}_yU_2^\dag U_2\hGR U_2^\dag U_2{\hat v}_yU_2^\dag U_2\hGA U_2^\dag\right]\\
&=\Sp_{\vec p}\left\{
\left[\frac{{\hat p}_y}m\hgr\frac{{\hat p}_y}m\hga\right]\sbarra_{{\hat p}_y\to{\hat p}_y-2mb}+
\left[\frac{{\hat p}_y}m\hgr\frac{{\hat p}_y}m\hga\right]\sbarra_{{\hat p}_y\to{\hat p}_y+2mb}\right\}\\
&=2\Sp_{\vec p}\left[\frac{{\hat p}_y}m\hgr\frac{{\hat p}_y}m\hga\right]=2\Sp_{\vec p}\left[\frac{{\hat p}_x}m\hgr\frac{{\hat p}_x}m\hga\right].
\end{split}\ee
Thus, we see that \cb{in the rotated CS $\sigma_{\alpha\alpha}$ are the same}
\cb{as in the absence of SOI for $a=\pm b$:}
\be
a=\pm b\quad\Longrightarrow\quad\sigma_{xx}=\sigma_{yy}=\sigma_{xx}(a=b=0).
\ee
% We conclude that, in case when $a=\pm b$, the \cb{the diagonal } conductivity tensor is the same as in the absence of SOI: $\sigma(a=\pm b)=\sigma(a=b=0)$.
% Note that this argument is not valid when we have a Zeeman term in the Hamiltonian.
This proves that the ZLA result (\ref{ZLAres}), as well as the
Boltzmann equation result\cite{JC:2006}, are incomplete.

\section{Further symmetries of $\sigma_{\alpha\beta}$\label{sec:os}}
While the Rashba SOI $V_R$ is invariant under rotations [see (\ref{invRashba})], the Dresselhaus SOI $V_D$ does not possess this symmetry --
rotation by an angle $\phi$ transforms it as follows:
\be \label{uno}
V_D\to{\vec e}_z\cdot\left[C\boldsymbol R_\phi\sigma\times R_\phi\hat{\vec p}\right],
\ee
which is in general different from
\be \label{due}
V_D={\vec e}_z\cdot\left[R_\phi C\boldsymbol\sigma\times R_\phi\hat{\vec p}\right]. %\quad\forall\phi.
\ee
In the special case of rotation by $\phi=\pi/2$, the anticommutator
\be
\left\{R_{\pi/2},C\right\}=\begin{pmatrix}0 & 0 &  0 \cr  0 & 0 & 0 \cr 0 & 0 & -2\end{pmatrix}
\ee
is zero in the $(x,y)$-plane, so that the r.h.s. of (\ref{uno}) differs from (\ref{due}) only by the sign.
On the other hand,
\be
R_{\pi/2}\begin{pmatrix}\sigma_{xx}&\sigma_{xy}\\ \sigma_{yx}&\sigma_{yy}\end{pmatrix}R_{\pi/2}^{-1}=
\begin{pmatrix}\sigma_{yy}&-\sigma_{yx}\\ -\sigma_{xy}&\sigma_{xx}\end{pmatrix},
\ee
so that rotation by $\pi/2$ is equivalent to the sign change of the Dresselhaus SOI amplitude $b$, and
\be\label{tre}
\sigma_{yy}(a,b)=\sigma_{xx}(a,-b),\quad\sigma_{xx}(a,b)=\sigma_{yy}(a,-b).
\ee
Even without the disorder averaging one can see that only  terms even in $(a,b)$ may produce a non-zero result of $\Sp_{\mathrm{spin}}$ in (\ref{ofodc}).
Together with the property (\ref{tre}) this leads to\cite{foooo}
\be \label{deltaSigmaxxyy}
\sigma_{yy}(-a,b)=\sigma_{xx}(a,b)=\sigma_{xx}(-a,-b)=\sigma_{yy}(a,-b).
\ee
%\msc{end of changes}

%\bibliography{refs.aps,books.aps,local.aps}

\begin{thebibliography}{26}
\bibitem[{exc()}]{exceptZLA}
\bibinfo{note}{Except for the ZLA, where the sum of the ``bubble'' and the
  ``vertex correction'' is given by a ``bubble'' where only one velocity
  operator is renormalized, see (\ref{ZLAres}).}
\bibitem[{foo({\natexlab{a}})}]{foot}
\bibinfo{note}{Thus in case of linear SOI, the classical (ZLA) conductivity
  tensor remains isotropic despite the anisotropy of the energy spectrum. This
  happens because the vertex renormalization cancels the anomalous part of the
  velocity operator. The same mechanism is also responsible for the vanishing
  of the spin-Hall conductivity (at ZLA)~\cite{aug04}.}
\bibitem[{foo({\natexlab{b}})}]{foooo}
\bibinfo{note}{Alternatively, (\ref{deltaSigmaxxyy}) follows from the fact that
  $\ts_1^*=-\ts_2$, $V_s$ is invariant under simultaneous complex conjugation,
  sign reversal of $ab$, and interchange $p_x\leftrightarrow p_y$.}
\end{thebibliography}
%\input{final-prb.bbl}

\end{document}